# Light-induced dipolar spectroscopy – A quantitative comparison between LiDEER and LaserIMD

Anna Bieber[1] , Dennis Bücker[1] , Malte Drescher*

Department of Chemistry and Konstanz Research School Chemical Biology, University of Konstanz, Konstanz, Germany

[1] These authors have contributed equally
* Corresponding author, malte.drescher@uni–konstanz.de

**Abstract**: Nanometric distance measurements with EPR spectroscopy yield crucial information on the structure and interactions of macromolecules in complex systems. The range of suitable spin labels for such measurements was recently expanded with a new class of light-inducible labels: the triplet state of porphyrins. Importantly, accurate distance measurements between a triplet label and a nitroxide have been reported with two distinct light-induced spectroscopy techniques, (light-induced) triplet-nitroxide DEER (LiDEER) and laser-induced magnetic dipole spectroscopy (LaserIMD). In this work, we set out to quantitatively compare the two techniques under equivalent conditions at Q band. Since we find that LiDEER using a rectangular pump pulse does not reach the high modulation depth that can be achieved with LaserIMD, we further explore the possibility of improving the LiDEER experiment with chirp inversion pulses. LiDEER employing a broadband pump pulse results in a drastic improvement of the modulation depth. The relative performance of chirp LiDEER and Laser-IMD in terms of modulation-to-noise ratio is found to depend on the dipolar evolution time: While LaserIMD yields higher modulation-to-noise ratios than LiDEER at short dipolar evolution times ($\tau = 2\,\mu s$), the high phase memory time of the triplet spins causes the situation to revert at $\tau = 6\,\mu s$.

## 1. Introduction

Pulsed dipolar EPR spectroscopy is a well-established technique to determine precise distance distributions between paramagnetic centers with distances ranging from around 1.6 nm up to 16 nm [1–3]. In combination with site-directed spin labeling, it is ideally suited for structural characterizations of macromolecules and complexes, and has emerged as valuable tool in structural biology [4–6]. Importantly, distance constraints can be obtained regardless of the overall size of the studied molecules, and measurements can be performed in the presence of membranes and even in cells [7–10]. Because the distance information is not encoded in signal intensity, but in the frequency domain, the result of a dipolar spectroscopy experiment is not a mean interspin distance, but a precise distance distribution reflecting structurally heterogeneous populations of macromolecules [1].

The most commonly employed dipolar spectroscopy technique is four-pulse double electron-electron resonance (DEER) [11] on nitroxide spin labels or metal complexes [12]. A very interesting recent addition to the collection of spin labels which are suitable for pulsed dipolar spectroscopy is the triplet state of porphyrins [13]. These chromophores are diamagnetic and thus EPR-silent in their ground state $S_0$, but absorption of light at a suitable wavelength leads to excitation of a fraction of the molecules to the first excited singlet state $S_1$, from which the paramagnetic triplet state $T_1$ with the spin quantum number $S_T = 1$ can then be formed by intersystem crossing (ISC). Next to being switchable by light, another attractive feature of porphyrins is their occurrence as endogenous prosthetic group e. g. in heme proteins [14] and photosynthetic complexes [15]. A further important phenomenon in this regard is optical spin polarization (OSP), that is, the fact that the initial population of the three triplet sublevels with $m_{S,T} = -1, 0, 1$ in high external magnetic fields does not correspond to a Boltzmann distribution immediately after ISC [16]. In consequence, the EPR spectrum of a spin-polarized triplet state shows both emissive and absorptive parts, and the signal intensity is significantly enhanced compared to systems in thermal equilibrium. The decay of the triplet EPR signal over time is governed by two mechanisms: spin-lattice relaxation between the triplet sublevels to thermal equilibrium, as well as the overall depopulation of the triplet state through ISC, non-radiative decay or phosphorescence, bringing the molecule back to the singlet ground state $S_0$ [16]. For porphyrins at cryogenic temperatures, these relaxation processes happen within milliseconds [14].

The first study that employed a porphyrin for light-induced dipolar spectroscopy was published by Di Valentin et al. [13]. In the experiment, the conventional 4-pulse DEER sequence was preceded by a laser flash to populate the triplet state in situ, which was then used as observer spin species, while the pump pulse was applied to the nitroxide. The pulse sequence of this technique, which will be called light-induced double electron-electron resonance (LiDEER) in the following, is schematically represented in Figure 1 (a). Potential advantages of LiDEER include a high observer signal intensity owing to OSP as well as clear spectral separation of the observed triplet from the pumped nitroxide [13].

A different approach for measuring the distance between a nitroxide and a porphyrin triplet state by light-induced dipolar spectroscopy was recently introduced by Hintze et al. [14]. In laser-induced magnetic dipole spectroscopy (LaserIMD), the most distinctive feature of the triplet state, i.e., the fact that its formation *in situ* can be controlled with high temporal precision with a laser pulse, is actively exploited in the pulse sequence. The basic idea is that instead of *changing* the dipolar interaction with a microwave inversion pulse during the echo sequence as it is done in (Li)DEER, the dipole-dipole coupling is *introduced* by triplet excitation at variable times during the observer pulse sequence [14]. The LaserIMD experiment is performed in practice by applying a Hahn echo sequence to the nitroxide, recording the primary spin echo and incrementing the position of this whole sequence relative to the (constant) position of a laser flash, as depicted in Figure 1 (b). It was shown in the original LaserIMD publication that as soon as the population of the triplet state with the laser flash occurs within the microwave pulse sequence at a time $t$ before the refocusing of the nitroxide spin echo, the dipolar interaction of the nitroxide and triplet spins during this time $t$ leads to a phase offset $\Delta\varphi_{dd} = m_{S,T}\omega_{dd}t$ at the time of echo formation if exchange coupling is neglected. The echo signal is thus modulated with $\cos(m_{S,T}\omega_{dd}t) = \cos(\omega_{dd}t)$ for $|m_{S,T}| = 1$, and the nitroxide-triplet distance distribution is extracted from the resulting form factor after background

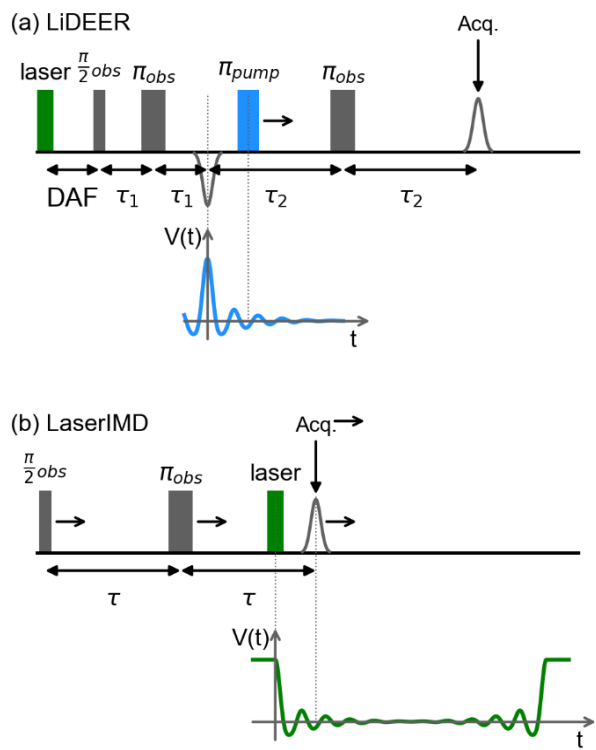

correction in an equivalent procedure to conventional DEER data [14,17].

The considerations outlined above are not only valid for the case where the laser flash occurs during echo refocusing, which is termed the "forward trace". Analogous considerations apply for the "reverse trace", where the laser flash comes between the two observer pulses, and the full LaserIMD trace thus has an approximately symmetric shape, where the reverse trace mirrors the forward trace [14].

Compared to nitroxide-nitroxide DEER, an increase in both sensitivity and modulation depth is expected for LaserIMD: Only one microwave frequency is needed which can be set to the center of a critically coupled resonator, yielding short observer pulses that are applied to the maximum of the nitroxide spectrum. Overall, this should maximize the number of excited observer spins, and thus the signal-to-noise ratio (SNR). Moreover, a temporal overlap of the laser flash with the microwave pulses or the acquisition trigger is not a problem, which allows the acquisition to be set on a primary echo, rather than the refocused echo as in the dead-time free DEER sequence. This leads to a further enhancement in SNR, since signal losses due to incomplete excitation by the refocusing pulse as well as transverse relaxation during the longer pulse sequence for a refocused echo are reduced. The modulation depth on the other hand only depends on the triplet quantum yield and the OSP, and can therefore be maximized by optimizing the conditions for triplet excitation [14].

In principle, for a porphyrin-nitroxide pair, both LiDEER and LaserIMD have been demonstrated to yield precise distance distributions [14,18]. As the attractive features of triplet-forming chromophores make these switchable spin labels very promising for future applications of pulsed EPR spectroscopy, the question that immediately arises is which of the two techniques would be preferable for studying a given system. In the present study, we set out to address this question by exploring the performance of LiDEER and LaserIMD under comparable conditions. To this end, we employ two of the spectroscopic rulers introduced recently by Di Valentin *et al.* as model compounds, with predicted distances of 2.3 nm (**1**) and 3.8 nm (**2**), respectively [18]. The structures of the model peptides are given in Figure 2 (a). Both LiDEER and

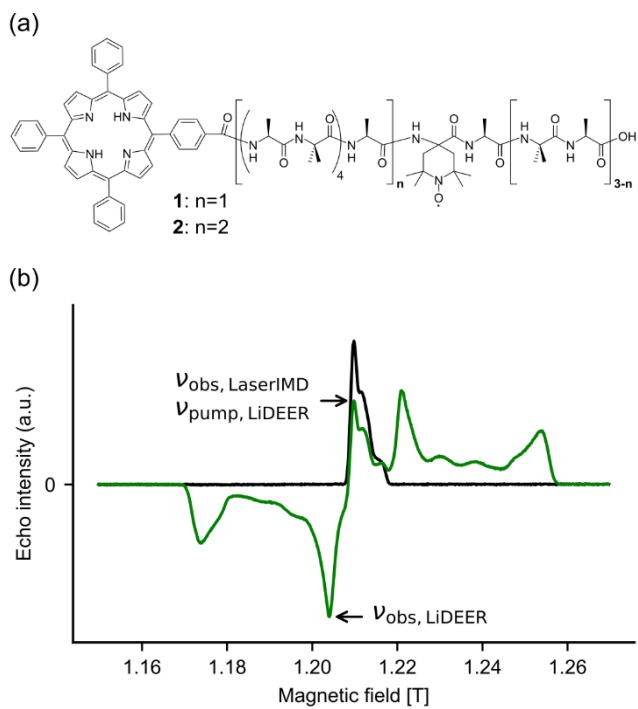

(a)

**1**: n=1
**2**: n=2

(b)

**Figure 2** (**a**) Structure of the model peptides **1** and **2**. The expected distance between the TOAC nitroxide spin label and the chromophore TPP is 2.3 nm for **1** and 3.8 nm for **2**, respectively. (**b**) Echo-detected field-swept spectra of 1 in Q band, with (green) and without (black) preceding laser irradiation (Delay after flash DAF = 250 ns). The spectral positions of observer and pump pulses for LiDEER and LaserIMD are indicated. Pulse settings: $\frac{\pi}{2}$ / $\pi$: 12/24 ns, $\tau$ = 1 µs.

LaserIMD were performed at Q band and with an optimized triplet excitation wavelength using a tunable laser system on identical samples to ensure maximal comparability. The resulting traces were analyzed with regard to the respective modulation-to-noise ratio. Furthermore, we demonstrate that in analogy to published results for DEER on other spin systems [19,20], the modulation depth of triplet-nitroxide LiDEER is enhanced dramatically when the rectangular pump pulse is replaced by a fast-passage chirp pulse.

## 2. Materials and Methods

For the comparison of LiDEER and LaserIMD, the spectroscopic rulers introduced by Di Valentin *et al.* for X-band LiDEER constitute ideal model systems. [18]. An N-terminal 5-(4-carboxyphenyl)-10,15,20-triphenylporphyrin (TPP) moiety serves as triplet label, and the nitroxide radical is introduced into the peptides by incorporation of a 4-amino-1-oxyl-2,2,6,6-tetramethylpiperidine-4-carboxylic acid (TOAC) moiety into the sequence of alternating L-alanine/α-aminoisobutyric acid (Ala/Aib) residues, which ensure an α-helical conformation of the peptides [13,21]. Figure 2 (a) shows the model peptides employed in the present work, with predicted interspin distances of 2.3 nm for peptide **1** and 3.8 nm for peptide **2** [18]. For EPR experiments, the model peptides **1** and **2** were dissolved in deuterated methanol with 2% $D_2O$ to a final concentration of 0.1 mM.

All EPR experiments were performed at Q band on a Bruker Elexsys E580 spectrometer equipped with a SpinJet-AWG unit and a 150 W pulsed TWT amplifier. Light excitation was performed at a wavelength of 515 nm.

Since the tunable laser system employed in the experiments allowed the use of a repetition rate of 50 Hz, all experiments were performed at a temperature of 30 K, with a shot repetition time of 20 ms. For the LaserIMD and LiDEER measurements on peptide **1**, an effective number of 128 shots per point (SPP) was accumulated with a dipolar evolution time of 2 µs, for peptide **2**, 384 SPP were accumulated with $\tau_{(2)}$ = 6 µs. The total number of accumulated SPP includes an 8-step phase cycle [22] combined with an 8-step nuclear modulation averaging procedure for LiDEER, and a 2-step phase cycle on the first observer pulse for LaserIMD, as described in detail in the SI.

For LiDEER measurements, the microwave pulse sequence was applied after the initial laser flash and a fixed delay time of 250 ns. Observer pulses were applied to the largest emissive peak in the triplet spectrum (see Figure 2), and the pump pulse was set to the center of the nitroxide peak, resulting in a frequency offset of 160 MHz between the pump and observer frequency. The resonator was fully overcoupled, and the positions of the observer and pump frequency with respect to the resonator profile were optimized to maximize the modulation-to-noise ratio (see Figure S4). For LiDEER with broadband inversion, a 120 ns chirp pulse with a (linear) frequency sweep over 100 MHz was used as pump pulse [19].

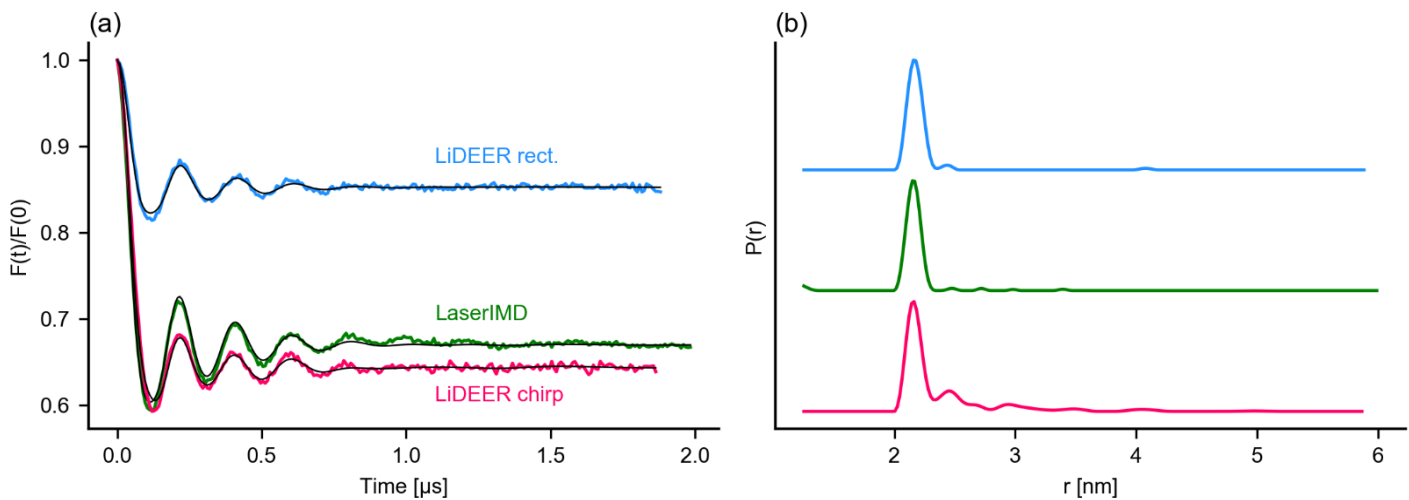

**Figure 3** Background corrected normalized dipolar evolution traces with respective fits (**a**) and corresponding distance distributions (**b**) for the short peptide **1**, measured with different techniques: LaserIMD is shown in green, LiDEER in blue (rectangular pump pulse) and magenta (100 MHz chirp pump pulse). (**a**) Dipolar evolution traces were measured with $\tau = 2$ μs. The LiDEER traces are shorter than the LaserIMD trace since an overlap of the pump pulse with the third observer pulse needed to be avoided. (**b**) Distance distributions obtained from the form factors in (a) via Tikhonov regularization, with $\alpha$ parameters chosen according to the L curve criterion. Mean distance $\langle r \rangle = 2.17$ nm and width $\sigma(r) = 0.06$ nm for LiDEER rect. and chirp and $\langle r \rangle = 2.16$ nm $\sigma(r) = 0.06$ nm for LaserIMD.

LaserIMD experiments were conducted in a critically coupled resonator (compare Figure S3), with the microwave frequency set to the center of the resonator profile.

The analysis of the LiDEER and LaserIMD data was performed with the software package DeerAnalysis2016 [17]. For LaserIMD, only the evaluation of the forward trace is shown throughout this study. After correcting for a three-dimensional homogeneous background, distance distributions were obtained from model-free fits of the form factors *via* Tikhonov regularization.

In pulsed dipolar spectroscopy, the crucial parameter when evaluating the performance of an experiment is the modulation-to-noise ratio (MNR), which is calculated as ratio between the modulation depth $\lambda$ and the noise in the normalized raw data. As a way to estimate this noise, similarly to published procedures [23], we used the standard error resulting from a seventh-order polynomial fit of the last part (see section 2 of SI) of the respective normalized time trace, where the modulation has already subsided.

For a more detailed description of Material and Methods as well as raw data see Supporting Information.

## 3. Results and Discussion

Echo-detected field-swept (EDFS) spectra of peptide **1** at Q band with and without light excitation are shown in Figure 2 (b). Without light excitation, the only paramagnetic species present is the TOAC nitroxide radical, and the spectrum shows the typical shape of a nitroxide spectrum with its maximum at a field of $B_0 = 1.21$ T. Upon excitation at 515 nm, the nitroxide spectrum is overlayed with the much broader spectrum of the TPP triplet state ($1.17 - 1.26$ T), which shows both emissive and absorptive peaks due to OSP [16]. As indicated in Figure 2 (b), the LiDEER observer pulses were always applied to the emissive extremum of the triplet spectrum, while the LiDEER pump pulse as well as the LaserIMD observer pulses were set to address the maximum of the nitroxide peak.

To allow a quantitative comparison of the two techniques, we recorded LiDEER and LaserIMD traces of one sample of peptide **1** at 30 K, keeping the laser settings ($\lambda = 515$ nm), repetition rate (50 Hz), dipolar evolution time ($\tau = 2$ μs) and effective SPP (128) constant throughout all experiments to ensure comparability. With these settings, the accumulation time for each experiment was less than 30 min. The resulting background-corrected LiDEER (blue) and LaserIMD (green) traces are shown in Figure 3 (a), with the respective distance distributions depicted in panel (b) of the figure. Already by inspecting the form factors (Fig. 3 (a)), it is evident that the LaserIMD trace (green) shows both a significantly higher modulation depth as well as a reduced noise level compared to LiDEER with a rectangular pump pulse (blue). This finding is reflected in the modulation depth, noise and MNR values derived from the data, which are summarized in Table 1. Importantly, the distance distributions derived *via* Tikhonov regularization do not significantly differ for LiDEER ($\langle r \rangle = 2.17$ nm, $\sigma = 0.06$ nm) and LaserIMD ($\langle r \rangle = 2.16$ nm, $\sigma = 0.06$ nm) (see Fig. 3(b)).

The outcome of this first comparison is easily understood by reflecting which experimental parameters determine the modulation depth and noise in LiDEER and LaserIMD, as the dependencies are actually quite different for both techniques.

The modulation depth $\lambda$ in LaserIMD depends on the triplet quantum yield $\Phi_T$ and the populations $p_{-1}$ and $p_{+1}$ of the triplet sublevels with $|m_{S,T}| = 1$ in the external magnetic field with [14]

$$\lambda = \Phi_T(p_{-1} + p_{+1}) \qquad (1).$$

Therefore, optimizing the triplet excitation can be expected to increase the modulation depth in LaserIMD. With the modulation depth of $\lambda = 34\%$ reached in the traces shown here, and assuming an equal population of the triplet sublevels with $m_{S,T} = -1,0,+1$ and thus $p_0 = \frac{1}{3}$ and $p_{-1} + p_{+1} = \frac{2}{3}$, the triplet quantum yield in our experiments can be estimated from equation (1) as $\Phi_T \approx 0.51$. This represents an improvement compared to the triplet quantum yield of $\Phi_T \approx 0.13$ reported by Hintze *et al.*, and can be explained by the optimized wavelength of 515 nm, compared to 351 nm in [14]. However, the triplet yield achieved here is still far from the predicted attainable triplet quantum yield of $\Phi_T \approx 0.9$ [14].

In contrast to LaserIMD, the modulation depth in LiDEER is entirely determined by the efficiency of the pump pulse. More precisely, the only molecules that are "visible" in the experiment are those where the porphyrin moiety was excited by the laser flash to the triplet state and then further addressed by the observer pulses, and the modulation depth $\lambda$ denotes the fraction of these molecules where the corresponding TOAC nitroxide spin (but not the triplet spin) is flipped by the pump pulse. Therefore, the best strategy for increasing $\lambda$ is to address as much of the entire nitroxide spectrum as possible without affecting the observed triplet spins.

In the LiDEER experiment under discussion, the rectangular pump pulse had a length of $t_p = 44$ ns and a corresponding excitation bandwidth (full width at half height) of $\Delta v_{1/2} \approx \frac{0.8}{t_p} \approx 18$ MHz [24]. With the nitroxide spectrum spanning more than 100 MHz, the room for improvement is evident. Frequency-swept pulses are well suited for this purpose, and the recent EPR literature brims with studies that demonstrate the tremendous gains in modulation depth that can be achieved by employing such broadband inversion pulses in DEER [19,24,25]. To see if we can achieve a similar enhancement in our experimental setting in LiDEER, we performed LiDEER using a fast-passage 120 ns chirp pump pulse with a bandwidth of 100 MHz, keeping all other experimental settings equal to the LiDEER experiment with the rectangular pump pulse.

The form factor and corresponding distance distribution of the LiDEER experiment performed with the 100 MHz chirp inversion pulse on peptide **1** are shown in magenta in Figure 3 (a) and (b), respectively. The improvement achieved with the chirp pump pulse is immediately evident in the form factor: The modulation depth of $\lambda = 36\%$ is more than twice the modulation depth of the experiment with a rectangular pump pulse, and even higher than LaserIMD (see Table 1). This finding is also reflected in the improved modulation-to-noise ratio of MNR=124 for chirp LiDEER. However, this MNR is still lower than for LaserIMD (MNR 207) due to the significantly higher noise level compared to LaserIMD.

Notably, the dipolar oscillations in the LiDEER chirp trace shows a reduced amplitude compared to the LaserIMD trace (Fig. 3 (a)). A dampening of the dipolar oscillation in DEER with adiabatic and fast-passage pump pulses has been explained by the fact that during the frequency sweep, spins with different resonance frequencies are inverted at different times. This results in an offset-dependent dispersion of the effective dipolar evolution times, since the zero time varies for spins at different spectral positions [25,26]. The magnitude of this dispersion is in the range of the pulse length of the frequency-swept pulse, and could thus account for the dampened oscillation observed when LiDEER is performed with a 120 ns chirp pump pulse. In contrast, the zero-time in LaserIMD is determined by the laser pulse (length 3.4 ns, jitter 0.3 ns). As a way to alleviate the problem for frequency-swept pump pulses in DEER, the use of the 5-pulse DEER sequence has been suggested [25], which does however come with the disadvantage that partial excitation artefacts need to be removed during data post-processing [27].

**Table 1** Modulation depth λ, noise (rmse) and modulation-to-noise ratio (MNR) of LiDEER and LaserIMD measurements on peptide **1**, with a dipolar evolution time of 2 µs and 128 (effective) SPP.

|  | λ [%] | rmse × 10³ | MNR |
|---|---|---|---|
| LiDEER rect. | 14.9 | 2.11 | 71.0 |
| LiDEER chirp | 35.9 | 2.90 | 124 |
| LaserIMD | 33.8 | 1.63 | 207 |

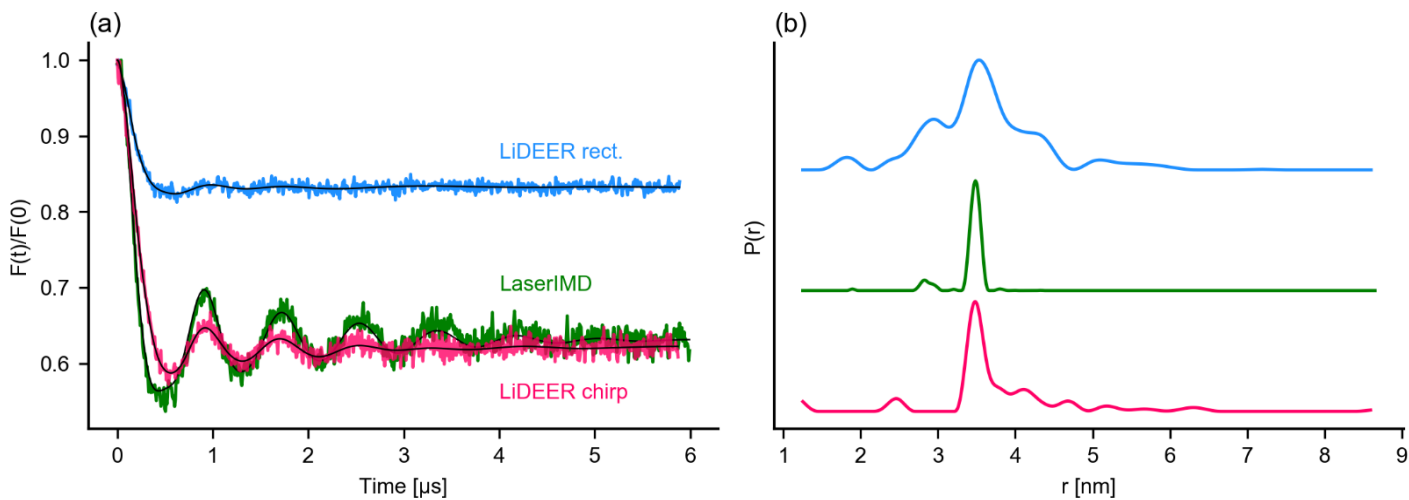

**Figure 4** Background corrected normalized dipolar evolution traces with respective fits (**a**) and corresponding distance distributions (**b**) for peptide **2**, measured with LaserIMD (green) and LiDEER with a rectangular pump pulse (blue) or a 100 MHz chirp pump pulse (magenta). (**a**) Dipolar evolution traces were measured with τ = 6 μs. The LiDEER traces are shorter than the LaserIMD trace since an overlap of the pump pulse with the third observer pulse needed to be avoided. (**b**) Distance distributions obtained from the form factors in (a) *via* Tikhonov regularization, with α parameters adjusted to obtain the fits indicated as black lines in (a). LiDEER rect.: mean distance ⟨r⟩ = 3.51 nm and width σ(r) = 0.50 nm; LaserIMD: ⟨r⟩ = 3.48 nm, σ(r) = 0.06 nm; LiDEER chirp: ⟨r⟩ = 3.53 nm, σ(r) = 0.14 nm.

Two factors can be identified that contribute to reducing the noise in LaserIMD compared to LiDEER: In contrast to LiDEER where two microwave frequencies with a 160 MHz offset need to be placed within the profile of an overcoupled resonator (see Figure S4), the observer frequency in LaserIMD is always chosen at the center of critically coupled resonator, which allows exploiting the highest available microwave field strength to maximize the excitation bandwidth of the observer pulses. The second advantage in LaserIMD is the fact that the signal can directly be observed on the primary echo, while the DEER signal is recorded on a refocused echo. If $\tau_2$ in LiDEER is identical to the interpulse delay of the primary echo sequence in LaserIMD, the overall evolution time is longer for the refocused echo, and even disregarding imperfect refocusing, a loss in signal intensity is expected due to transverse relaxation. Observing the refocused echo is necessary in

DEER in order to get a dead time-free trace, as a temporal overlap of the pump pulse with any of the observer pulses needs to be avoided. On the other hand, no technical restrictions prohibit an overlap of the laser flash with the microwave pulses (or the acquisition trigger) in LaserIMD, which is why the primary echo can be observed directly in this case [14].

Next to these technical differences between LaserIMD and LiDEER, another important point to note is the fact that different spin species are observed in the two experiments. For LiDEER, the number of observable (triplet) spins is determined by the light excitation and triplet quantum yield, while at the same time, a signal enhancement is expected due to OSP [13,16]. A further parameter that becomes crucial when measuring longer dipolar traces required for detecting longer distances [28] is the phase memory time of the observer spins. For the TPP triplet spins, the phase memory time at 30 K was determined as 6.0 μs, while the TOAC nitroxide spins which are observed in LaserIMD yielded $T_m = 1.8$ μs (see Figure S1).

The influence of these different phase memory times becomes apparent in the LaserIMD and LiDEER experiments performed on peptide **2**: In order to resolve the distance distribution around the expected value of 3.8 nm [18], the dipolar evolution time was increased from 2 μs for peptide **1** to 6 μs for the peptide **2**, which should yield a reliable distribution shape up to a distance of 4.3 nm [29]. Figure 4 shows the resulting form factors (a) and distance distributions (b) for LaserIMD and LiDEER with a rectangular and a 100 MHz chirp pump pulse, respectively, and the relevant values for a quantitative analysis are summarized in Table 2.

**Table 2** Modulation depth λ, noise (rmse) and modulation-to-noise ratio (MNR) of LiDEER and LaserIMD measurements on peptide **2**, with a dipolar evolution time of 6 µs and 384 (effective) SPP.

|  | λ [%] | rmse × $10^3$ | MNR |
|---|---|---|---|
| LiDEER rect. | 16.7 | 4.39 | 38.0 |
| LiDEER chirp | 37.9 | 6.89 | 55.0 |
| LaserIMD | 37.1 | 8.06 | 46.1 |

Similarly to the modulation depths obtained for peptide 1 (Fig. 3), for peptide 2, the modulation depth of LiDEER with a rectangular pump pulse reaches less than half of the values obtained for LaserIMD and LiDEER with a chirp pump pulse (Fig. 4 a). On the other hand, the noise level of the LaserIMD trace is no longer lower than for LiDEER, but even surpasses the noise in both LiDEER traces (see Table 2). This finding indicates that at a dipolar evolution time of 6 µs, the benefit of the single microwave frequency and primary echo observation in LaserIMD is counteracted by the shorter phase memory time of the employed nitroxide compared to the observed triplet spins in LiDEER. In consequence, it is now the LiDEER experiment with the chirp pump pulse that gives the highest MNR, followed by LaserIMD, with the classical LiDEER still yielding the lowest MNR due to the low modulation depth (see Table 2).

Of note, the modulation-to-noise ratio in LaserIMD could be enhanced by choosing a persistent spin label with a longer phase memory time at cryogenic temperatures than the TOAC nitroxide used here, such as a triarylmethyl (trityl) tag [30].

While for all three experiments a mean distance of 3.5 nm was determined, the respective widths of the distance distributions differ significantly between the measurements (Fig. 4 (b)). The low modulation depth in the trace of the LiDEER measurement with a rectangular pump pulse impedes the determination of an accurate fit at the given noise level and yields a much broader distance distribution than those obtained in the remaining two experiments. Even though the LaserIMD yields an even higher noise (see Table 2), in this case, the high modulation depth and large amplitude of the dipolar oscillation allow a reliable distance analysis.

## 4. Conclusion

In the comparison of the two recently introduced light-induced dipolar spectroscopy methods, we find that LaserIMD yields significantly higher modulation depths than LiDEER with rectangular pump pulses. However, this intrinsic disadvantage of LiDEER is alleviated with the use of a broadband inversion pulse for LiDEER. In this case, the signal-to-noise ratio becomes crucial for the comparison, and the relative performance of both techniques is found to depend on the chosen dipolar evolution time: For a short evolution time of 2 µs, LaserIMD yields a lower noise level, as it is performed with a primary echo sequence at a single microwave frequency. At higher evolution times, however, the longer phase memory time of the porphyrin triplet compared to the TOAC nitroxide results in an enhancement of the LiDEER signal compared to LaserIMD, and thus in a higher modulation-to-noise ratio for LiDEER if it is performed with a broadband pump pulse. Importantly, this implies that the LaserIMD signal can be enhanced with the use of spin labels with longer phase memory times than TOAC. In addition, given the dramatic improvement in modulation depth for LiDEER with a broadband pump pulse, it would certainly be worth it to investigate the impact of a larger set of inversion pulse shapes and parameters on the experiment. For LaserIMD, the modulation depth can be tuned *via* the triplet excitation, an enhancement of the SNR e. g. with optimal control theory-derived pulses as observer pulses can be envisioned [31,32].

In the end, both LaserIMD and LiDEER (with a broadband inversion pulse) are shown to yield excellent modulation-to-noise ratios within short accumulation times in this study. An important next step will be to identify and develop new triplet labels for both techniques and explore their combinations with varying stable spin labels in order to unravel the full potential of light-induced dipolar spectroscopy for future applications.

## Acknowledgements

This project has received funding from the European Research Council (ERC) under the European Union's Horizon 2020 research and innovation programme (Grant Agreement number: 772027 — SPICE — ERC-2017-COG).

A.B. gratefully acknowledges support by the German National Academic foundation. We thank Sonja Tischlik and Mykhailo Azarkh for helpful discussions.

# Supporting Information for:
# Light-induced dipolar spectroscopy – A quantitative comparison between LiDEER and LaserIMD

Anna Bieber[1], Dennis Bücker[1], Malte Drescher*

*Department of Chemistry and Konstanz Research School Chemical Biology, University of Konstanz, Konstanz, Germany*

[1] These authors have contributed equally

* Corresponding author, malte.drescher@uni–konstanz.de

# Contents

# 1 Extended Materials and Methods

## 1.1 Instrumentation

### 1.1.1 EPR spectrometer

All EPR experiments were performed at Q band (34 GHz) on a commercially available Bruker Elexsys E580 spectrometer operating with a SpinJet-AWG unit (Bruker Biospin, Rheinstetten, Germany) and a 150 W pulsed traveling-wave tube (TWT) amplifier (Applied Systems Engineering, Fort Worth, USA). Samples were held at cryogenic temperatures (30 K) with the EPR Flexline helium recirculation system (CE-FLEX-4K-0110, Bruker Biospin, ColdEdge Technologies) comprising a cold head (expander, SRDK-408D2) and a F-70H compressor (both SHI cryogenics, Tokyo, Japan), controlled by an Oxford Instruments Mercury ITC.

The commercial Q-band resonator (ER5106QT-2, Bruker Biospin) for 3 mm outer diameter sample tubes was overcoupled for experiments with two microwave (m.w.) frequencies (LiDEER), and critically coupled for single frequency experiments (LaserIMD).

### 1.1.2 Laser system

The tunable diode pumped Nd:YAG laser system NT230-50-ATTN2-FC (EKSPLA, Vilnius, Lithuania) comprising a pump laser, harmonics generators (SHG,THG) and an optical parametric oscillator (OPO) was used for light excitation at 515 nm. The system was operated with shot frequencies of 50 Hz and pulse energies optimized for each experiment ($5 - 10$ mJ). The average pulse length is 3.4 ns with a jitter of 0.3 ns. Triggering was performed with the pulse PatternJet of the EPR spectrometer, and the light was coupled into the resonator using a quartz glass fiber (1 mm core, Pigtail WF 1000 / 1100 / 1600 T, CeramOptec GmbH, Bonn, Germany) which was introduced at the top of the sample support, with the fiber end adjusted to a height of 5 mm above the sample surface.

## 1.2    Sample preparation

**Model peptide** The model peptides TPP-Ala-(Aib-Ala)$_4$-TOAC-(Aib-Ala)$_2$-OH (**1**) and TPP-Ala-(Aib-Ala)$_4$-Ala-Ala-(Aib-Ala)$_4$-TOAC-Ala-Aib-Ala-OH (**2**) were purchased from Biosyntan GmbH, Berlin, Germany. For EPR experiments, the peptides were dissolved in perdeuterated methanol with 2 vol% D$_2$O (both Sigma-Aldrich /Merck KGaA, Darmstadt, Germany) to a final concentration of 0.1 mM. The filling height was 3.5 mm in a 3 mm o.d. quartz glass tube, which was centered vertically in the resonator mode.

## 1.3    EPR experiments

All pulsed EPR experiments were performed at Q band (34 GHz) with a microwave attenuation of 0 dB at 30 K. The shot repetition time was chosen as 20 ms to avoid saturation of the relative longitudinal relaxation time. Echo signals were detected in integrator mode with a video bandwidth of 200 MHz and an integrator gate width corresponding to the respective $\pi$ pulse length.

Hahn echo sequences were always performed as $\pi/2 - \tau - \pi - \tau -$ echo, with an interpulse delay of $\tau = 1$ µs and pulse lengths optimized with a nutation experiment. Processing of the data was done in MATLAB R2018a [1] using the software packages easyspin (9.4.0) [2] and DeerAnalysis2016 [3].

All EPR experiments were performed at least two times to ensure reproducibility.

### 1.3.1   Standard experiments

**EDFS** Echo-detected field-swept spectra were recorded with a Hahn echo sequence, a sweep width of 1200 G and 10 shots per point. For echo-detected field sweeps with light excitation, the laser pulse was set to precede the m.w. pulse sequence with an approximate DAF of 250 ns.

**Echo decay measurements** Phase memory relaxation was measured by increasing the interpulse delay of a Hahn echo sequence, starting with $\tau = 1$ µs, and integrating the echo. The resulting trace was fit with a mono-exponential decay $V(t) = V(0) \cdot e^{-t/T_m}$ to extract the phase memory time $T_m$.

**Resonator profile** Resonator profiles were measured as described in the literature [4], by determining the m.w. field strength $\nu_1$ as nutation frequency at different m.w. frequencies. Nutation experiments were performed over a frequency window of 400 MHz in 10 MHz steps, and the magnetic field was co-stepped to retain correct observer pulse flip angles. The respective nutation frequencies were determined by Fourier transformation of the resulting nutation traces.

### 1.3.2 LiDEER

Light-induced DEER (LiDEER) measurements were performed in an overcoupled resonator, with the observer frequency set to the global minimum of the triplet spectrum and the pump pulse frequency (center frequency for chirp pump pulse) chosen at an offset of – 160 MHz from the observer frequency. Pulse lengths (Table S1) were always optimized for the respective frequencies and spin species with nutation experiments, and the magnetic field $B_0$ was set to the maximum of the nitroxide spectrum at the pump pulse frequency.

The laser flash (wavelength 515 nm, approximate pulse energy 5-10 mJ) was set to precede the 4-pulse DEER sequence by a delay after flash (DAF) of 250 ns. Nuclear modulation averaging was performed by incrementing the first observer interpulse delay in eight steps of 16 ns each from an initial value of 600 ns. As described by Tait and Stoll [5], the use of a coherent pump pulse introduces artefacts into the DEER trace which need to be removed by phase cycling. To this end, we employed one of their suggested 8-step phase cycles in LiDEER experiments, where a 2-step $[+(+x) - (-x)]$ phase cycle on the first observer pulse was combined with a 4-step $[+(+x) + (+y) + (-x) + (-y)]$ phase cycle on the pump pulse.

Chirp pump pulses were generated with the built-in function of the Bruker SpinJet-AWG unit and programmed as 120 ns linear frequency sweep from -210 MHz to -110 MHz offset from the observer frequency.

### 1.3.3   LaserIMD

LaserIMD measurements were performed in a critically coupled resonator, if not stated otherwise. Pulse lengths (Table S1) were always optimized with nutation experiments. A 2-step $[+(+x) - (-x)]$ phase cycle was applied on the first observer pulse to cancel receiver offsets. The full LaserIMD with forward and reverse trace was recorded, though only the forward trace was used for the evaluations shown in the present study. All measurements were recorded as two-dimensional dataset with 8 shots per point. During post-processing, the phase in each scan was corrected individually, and phase-corrected scans were added to yield the accumulated raw data with the same effective number of accumulations per point as in the LiDEER measurements of the same sample.

**Table S1** Overview of the pulse parameters for the LiDEER and LaserIMD experiments of peptide **1** and **2**.

|  | LiDEER | | | LaserIMD |
|---|---|---|---|---|
|  | $\pi_{obs}/2 : \pi_{obs}$ / ns | $\pi_{pump}$ (rect.) / ns | $\pi_{pump}$ (chirp) / ns | $\pi_{obs}/2 : \pi_{obs}$ / ns |
| Peptide **1** | $19 : 38$ | 44 (rect.) | 120 (chirp) | $12 : 24$ |
| Peptide **2** | $21 : 42$ | 34 (rect.) | 120 (chirp) | $12 : 24$ |

# 2 Analysis of distance measurements

## 2.1 Distance Analysis

The evaluation of LiDEER and LaserIMD data was performed with DeerAnalysis2016 (04.12.2017) [3], using only the forward trace in the case of LaserIMD. The zero time was determined manually, and starting times for background fits (three-dimensional homogeneous background) were determined automatically by the program.

The output command for the L-curve-file in DeerAnalysis2016 in save_result.m, line 131 was changed to:

```
data5=[handles.Lcurve_rho handles.Lcurve_eta handles.regpars]
```

## 2.2 Determination of MNR-ratio

The modulation-to-noise (MNR)-ratio was determined as

$$\text{MNR} = \frac{\text{modulation depth } \lambda}{\text{noise}}$$

The modulation depth was determined by DeerAnalysis and was used directly. The noise level was extracted from the last parts of the traces. For the short peptide **1** the range of 1 to 1.8 µs was used, for the longer peptide **2** the range of 4 to 5.8 µs was used. This part of the normalized form factor was fitted by a $7^{\text{th}}$ order polynomial function. The root mean squared error (rmse) of this fit was used as noise level.

# 3     Phase memory time

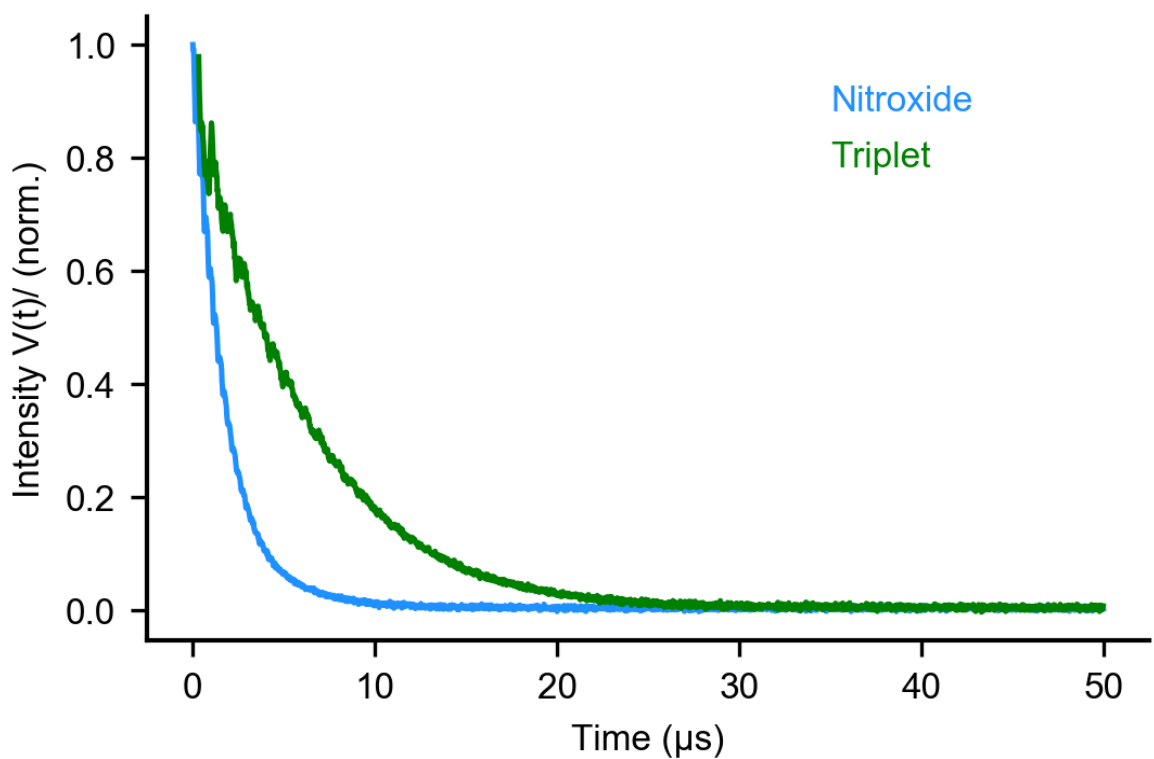

**Figure S1** Echo relaxation measurement at the spectral position of the nitroxide at 1.2 T (blue) and the minimum of the triplet at 12.04 T (green). The first 15 data points for the Triplet measurement were neglected, due to laser power build-up time. Phase memory times determined with monoexponential fits are 1799.4 ns (nitroxide) and 6027.7 ns (triplet).

# 4      Echo-detected-field-sweep of peptide 2

The EDFS of peptide **2** (Figure S2) shows no significant difference with respect to peptide **1**. The same spectral positions were used for the respective pump and observer pulses for both peptides.

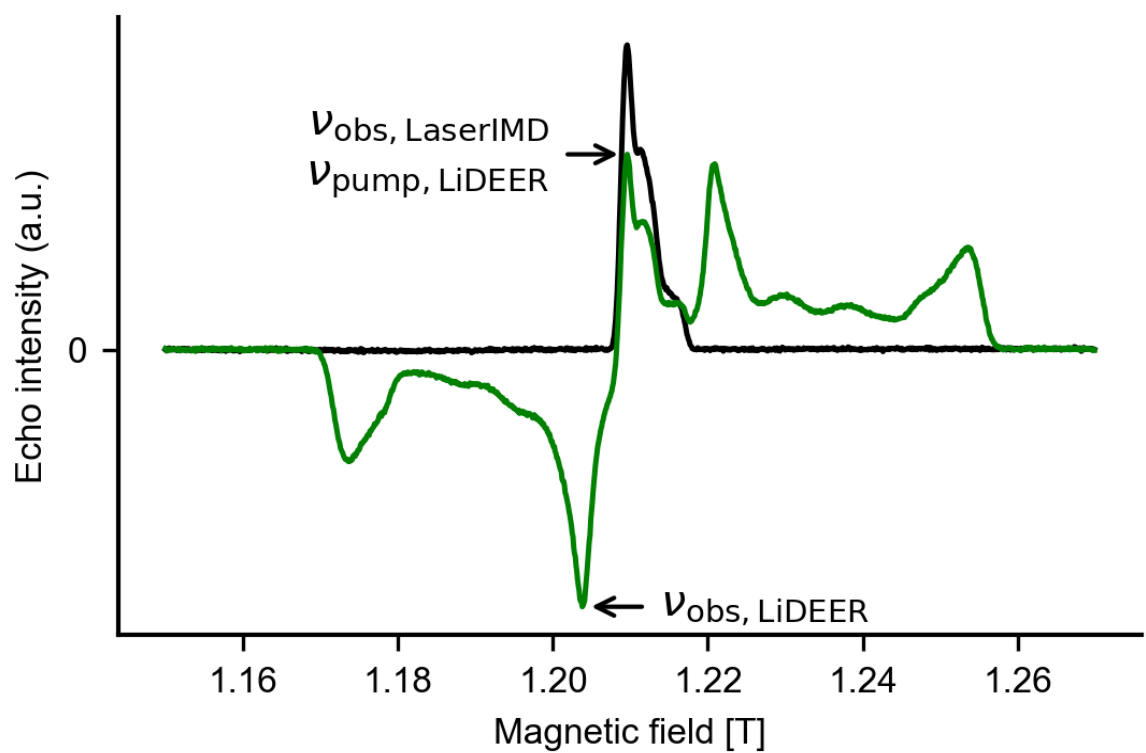

**Figure S2** Echo-detected-field-sweep of peptide **2** with (green) and without (black) laser irradiation. Pulse settings were 12/24 ns for Hahn echo with a τ=1000 ns. For measurement with laser irradiation DAF was 250 ns. Spectral pump position for LiDEER and spectral observing positions for LaserIMD and LiDEER are indicated.

# 5    LaserIMD with overcoupled resonator

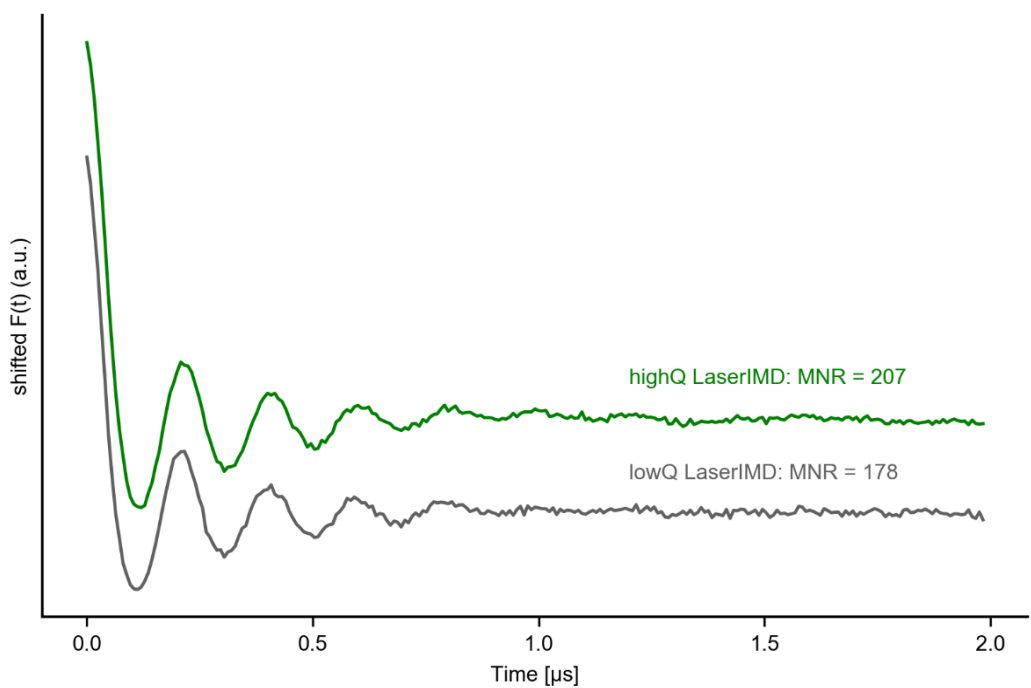

**Figure S3** Forward LaserIMD time traces of peptide **1** in a critically coupled (highQ, green) and an overcoupled (lowQ, grey) resonator. Modulation depth of 33.9 % for the highQ case yield a MNR(highQ) = 207 and 33.4 % for the lowQ case a MNR(lowQ) = 178. Pulse parameters for the lowQ LaserIMD were 16/32 ns.

To demonstrate the positive effect of measuring LaserIMD in a critically coupled resonator, the experiment was repeated in an overcoupled resonator for peptide **1** (Figure S3). Both measurements show nearly the same modulation depth and differ only in the noise level as expected (critically coupled: $1.634 \cdot 10^{-3}$, overcoupled: $1.878 \cdot 10^{-3}$). In our case, a critically resonator yields a 16 % better MNR ratio.

# 6     LiDEER with asymmetric pump/observer position

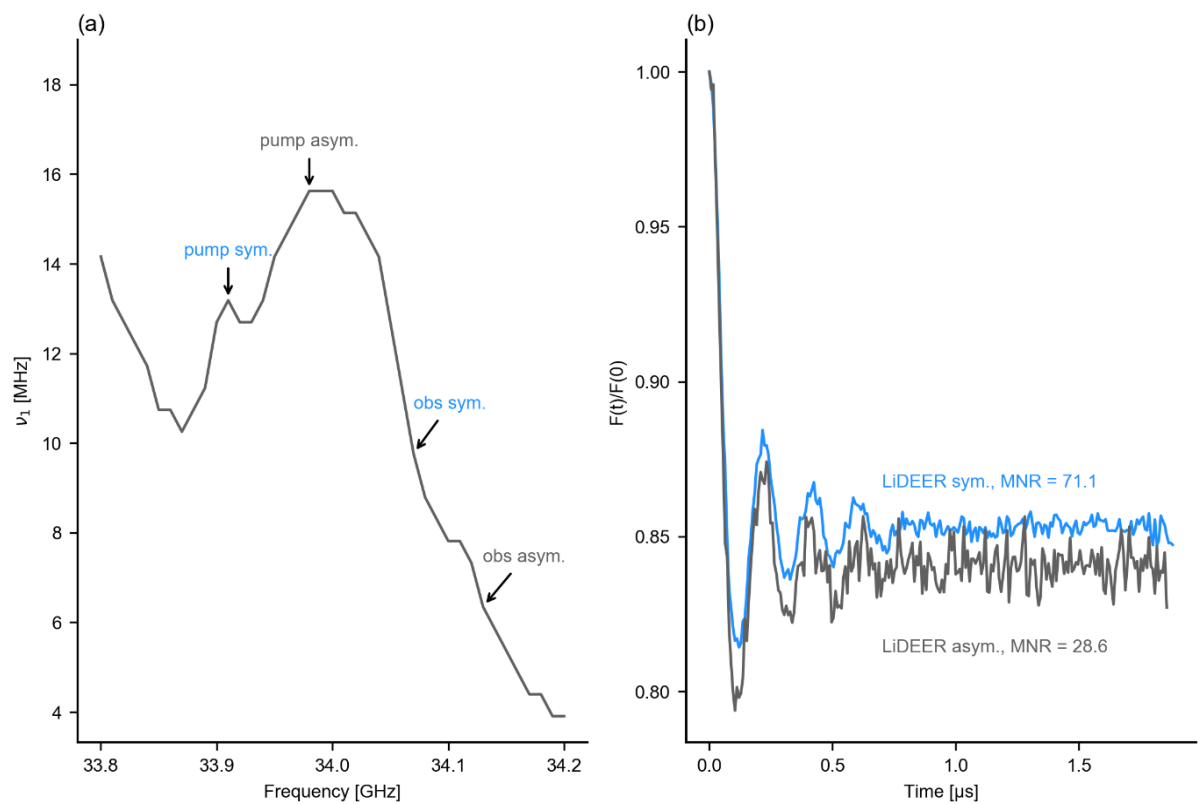

**Figure S4** Comparison of LiDEER (peptide **1**) with symmetric and asymmetric pump/observer pulse positions. (**a**) Experimentally determined resonator profile of the overcoupled resonator with symmetric (blue) and asymmetric (black) pulse positions, depicted with arrows. (**b**) Normalized form factors for symmetric (blue) and asymmetric (black) pulse positions. The asymmetric (black) pulse positions gives a higher modulation depth (15.9 %) but a lower MNR (28.6) than the symmetric (blue) pulse position with 15.0% modulation depth, but a higher MNR (71.1)

In the case of LiDEER, a symmetric positioning of the pulses around the center frequency of the resonator is favorable to obtain the best MNR for our setup (Figure S4). While LiDEER yields a higher modulation depth when the pump pulse is positioned at the maximum of the resonator profile, the MNR is significantly higher with a symmetrical positioning around the maximum. This can be attributed to the decrease in noise level for asymmetric positioning (symmetric: $2.108 \cdot 10^{-3}$, asymmetric: $5.573 \cdot 10^{-3}$). With a symmetric positioning of the pulses a gain in MNR of a factor 2.5 is achieved.

# 7 Pulse excitation profiles for LiDEER

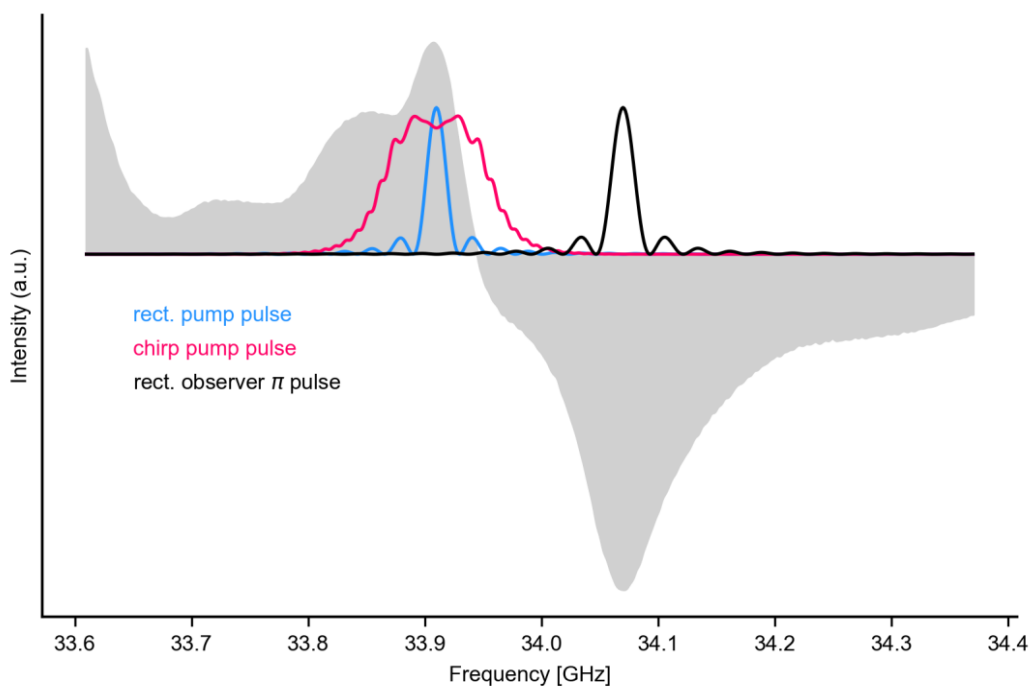

**Figure S5** Calculated excitation profiles for the pulses in the LiDEER experiments on peptide **1**. In grey, the shape of a laser-induced EDFS is shown on a frequency axis. The maximum of the nitroxide signal corresponds to 33.91 GHz and the minimum (absolute maximum) of the triplet spectrum to 34.07 GHz. This results in a frequency offset $\Delta \nu$ = -160 MHz. A rectangular observer $\pi$ pulse (38 ns) is shown in black, the corresponding rectangular pump pulse (44 ns) in blue. In magenta, the respective chirped pump pulse ($\Delta f$ = 100 MHz, 120 ns) is shown.

# 8 Evaluation of distance distributions from LaserIMD and LiDEER experiments: full datasets

## 8.1 Peptide 1

### 8.1.1 LaserIMD

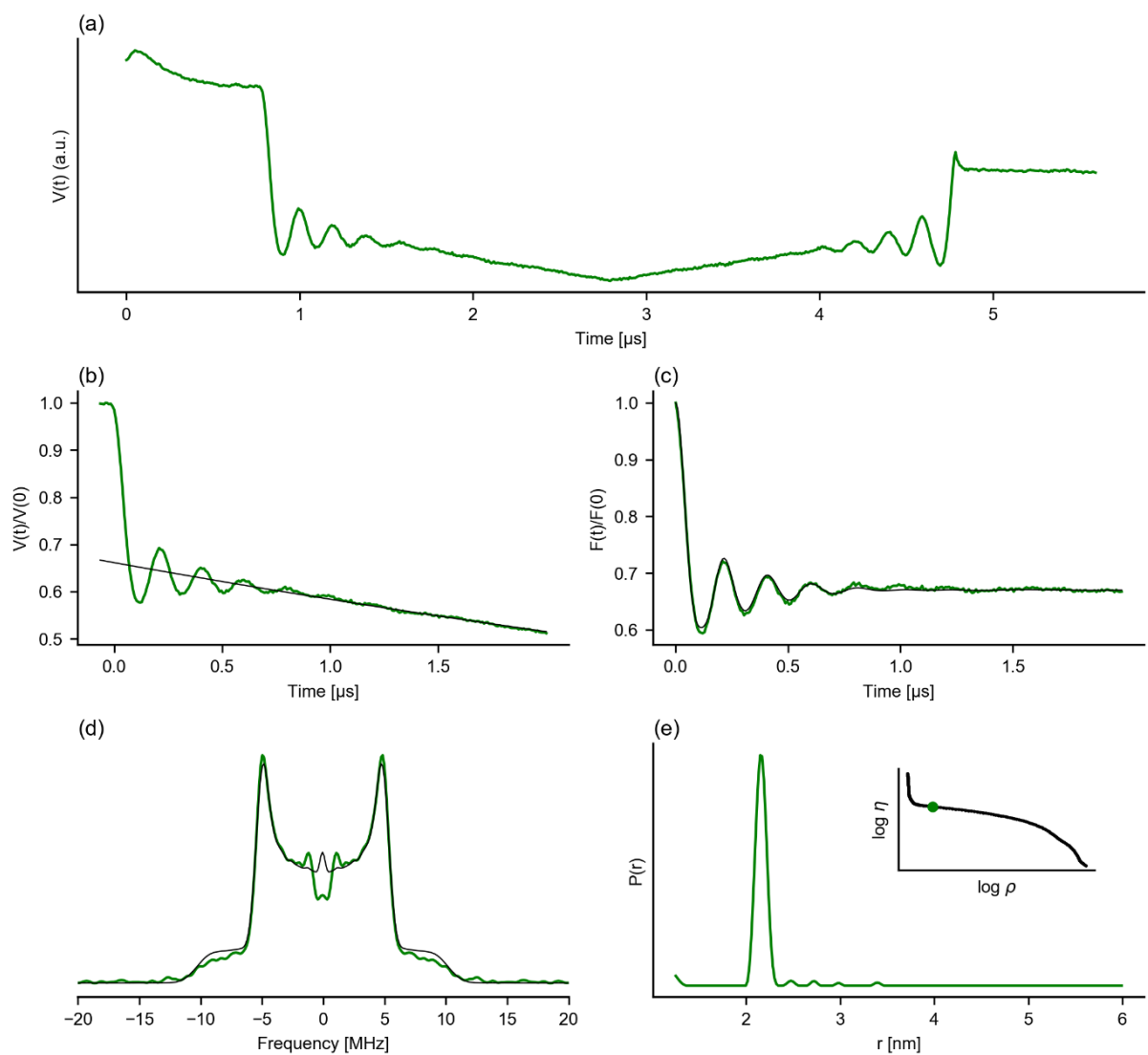

**Figure S6** Distance measurement with LaserIMD of peptide **1**. (**a**) Full LaserIMD trace with forward and reverse part. (**b**) Forward part of the normalized LaserIMD time trace with three-dimensional background fit. Zero time 784 ns, automatically determined background start 336 ns, cut-off at 1984 ns. (**c**) Background corrected form factor, modulation depth 33.9 %, with fit by Tikhonov regularization. (**d**) Dipolar spectrum obtained by Fourier transformation and corresponding fit. (**e**) Distance distribution obtained by Tikhonov regularization, $\langle r \rangle$ = 2.16 nm, width $\langle s \rangle$ = 0.06 nm. The L curve is shown as inset, the chosen $\alpha$-parameter of 3.2 is indicated.

### 8.1.2 LiDEER rectangular pump

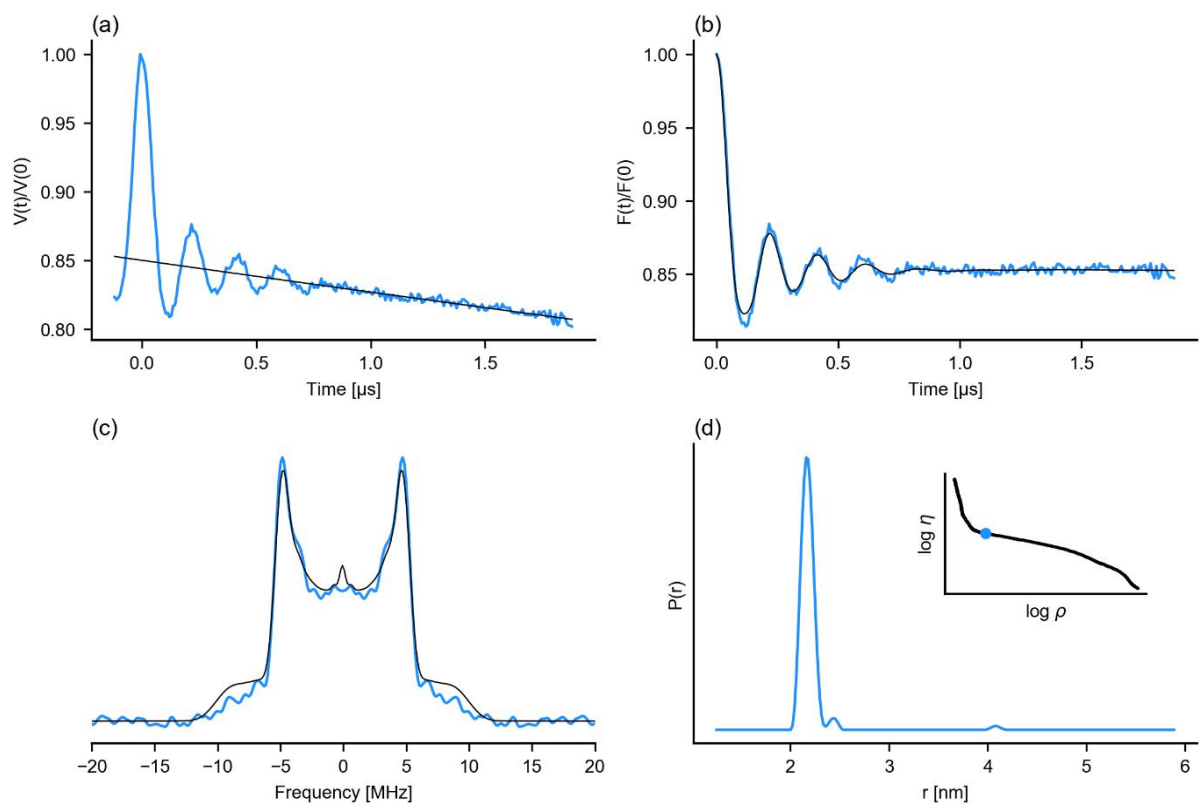

**Figure S7** Distance measurement with LiDEER (rectangular pump pulse) of peptide **1**. (**a**) Normalized time trace with three-dimensional background fit. Zero time 427 ns, automatically determined background start 896 ns, cut-off at 1881 ns. (**b**) Background corrected form factor, modulation depth of 15.0 %, with fit by Tikhonov regularization. (**c**) Dipolar spectrum obtained by Fourier transformation and corresponding fit. (**d**) Distance distribution obtained by Tikhonov regularization, $\langle r \rangle$ = 2.17 nm, width $\langle s \rangle$ = 0.06 nm. The L curve is shown as inset, chosen $\alpha$-parameter of 5.7 is indicated.

### 8.1.3 LiDEER chirped pump

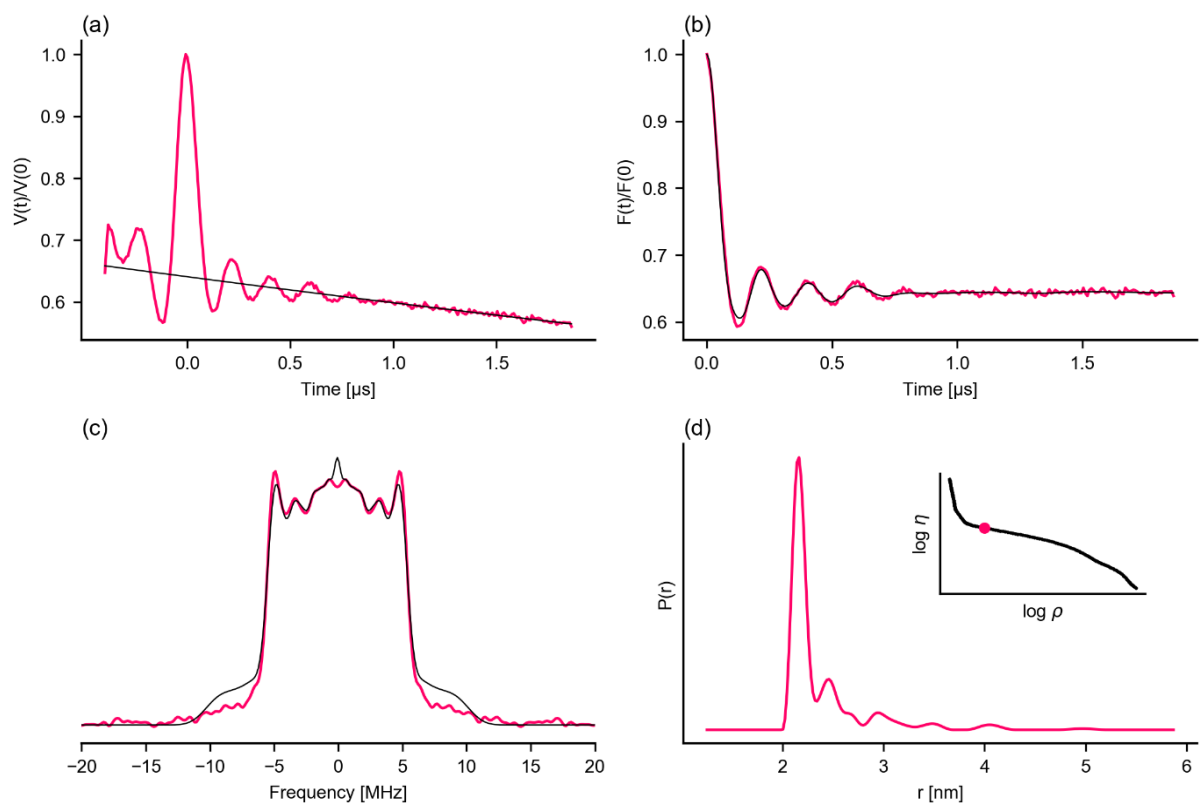

**Figure S8** Distance measurement with LiDEER (chirped pump pulse) of peptide **1**. (**a**) Normalized time trace with three-dimensional background fit. Zero time of 403 ns, automatically determined background start 344 ns, cut-off at 1864 ns. (**b**) Background corrected form factor, modulation depth 35.9 %, with fit by Tikhonov regularization. (**c**) Dipolar spectrum obtained by Fourier transformation and corresponding fit. (**d**) Distance distribution obtained by Tikhonov regularization, ⟨r⟩ = 2.17 nm, width ⟨s⟩ = 0.06 nm. The L curve is shown as inset, the chosen α-parameter of 5.6 is indicated.

## 8.2 Peptide 2

### 8.2.1 LaserIMD

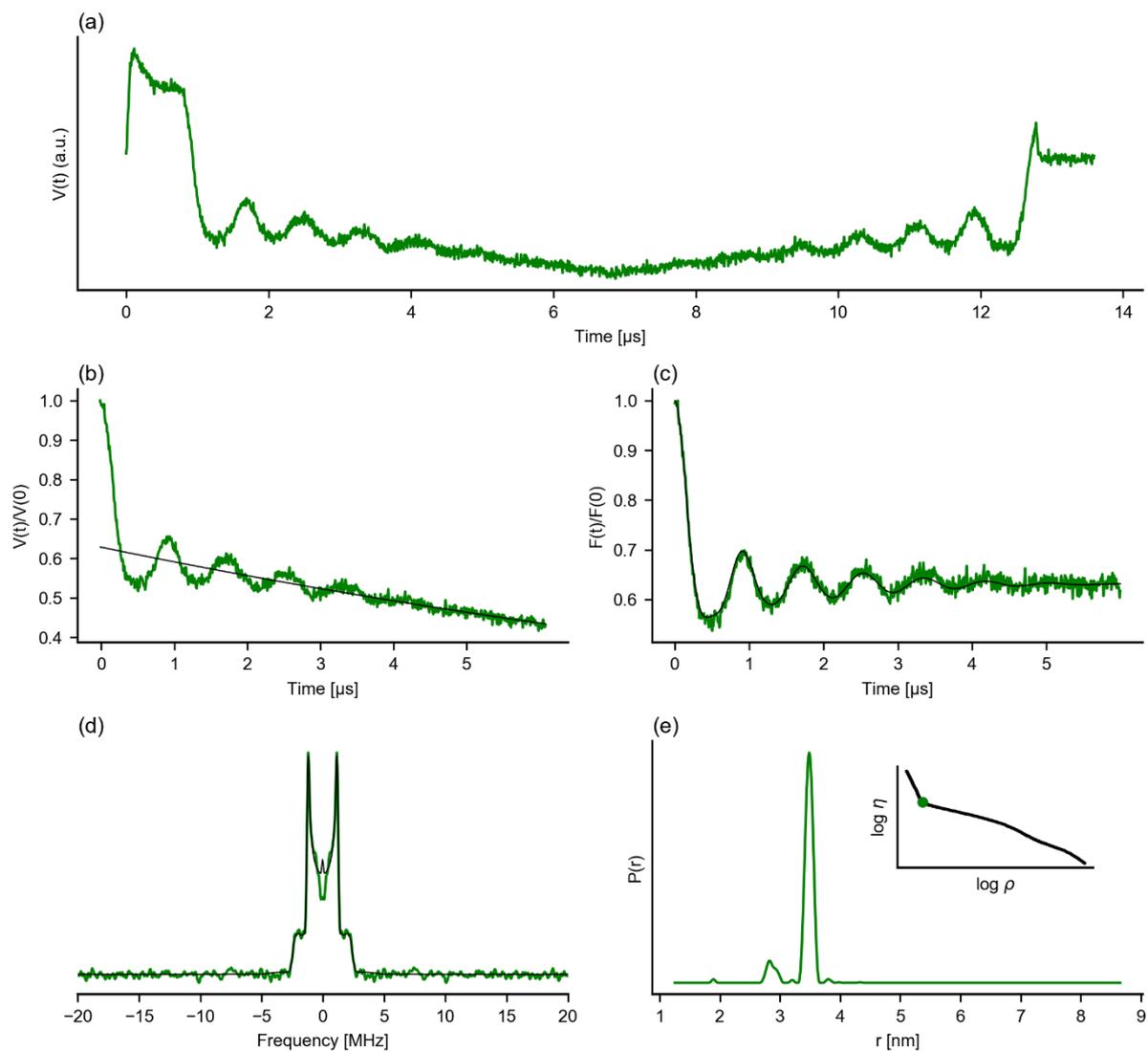

**Figure S9** Distance measurement with LaserIMD of peptide **2**. (**a**) Full LaserIMD trace with forward and reverse part. (**b**) Forward part of the normalized LaserIMD time trace with three-dimensional background fit. Zero time 772 ns, automatically determined background start 1328 ns, cut-off at 5984 ns. (**c**) Background corrected form factor, modulation depth 37.2 %, with fit by Tikhonov regularization. (**d**) Dipolar spectrum obtained by Fourier transformation and corresponding fit. (**e**) Distance distribution obtained by Tikhonov regularization, $\langle r \rangle$ = 3.48 nm and width $\langle s \rangle$ = 0.06 nm. The L curve is shown as inset, the chosen α-parameter of 58 is indicated.

## 8.2.2 LiDEER rectangular pump

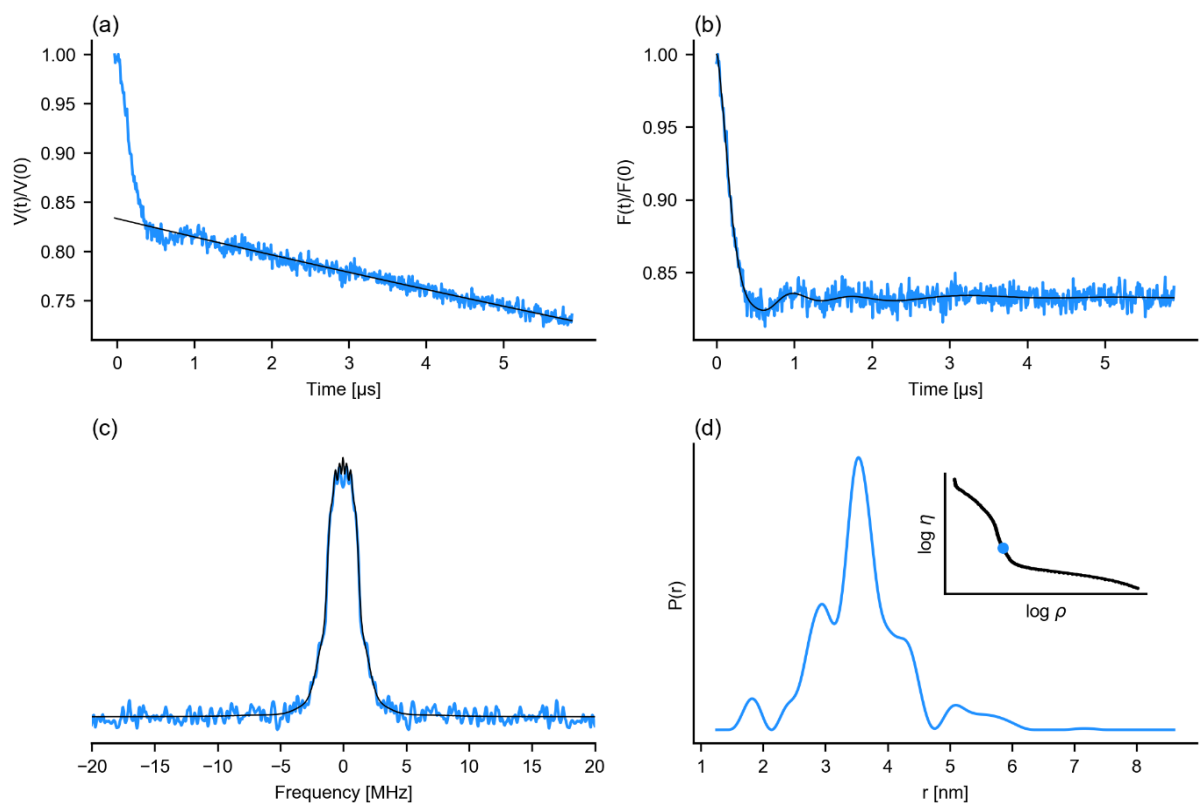

**Figure S10** Distance measurement with LiDEER (rectangular pump pulse) of peptide **2**. (**a**) Normalized time trace with three-dimensional background fit. Zero time 420 ns, automatically determined background start 2312 ns, cut-off at 5892 ns. (**b**) Background corrected form factor, modulation depth 16.7 %, with fit by Tikhonov regularization. (**c**) Dipolar spectrum obtained by Fourier transformation and corresponding fit. (**d**) Distance distribution obtained by Tikhonov regularization, $\langle r \rangle$ = 3.51 nm and width $\langle s \rangle$ = 0.50 nm . The L curve is shown as inset, the chosen α-parameter of 829.1 is indicated.

### 8.2.3 LiDEER chirped pump

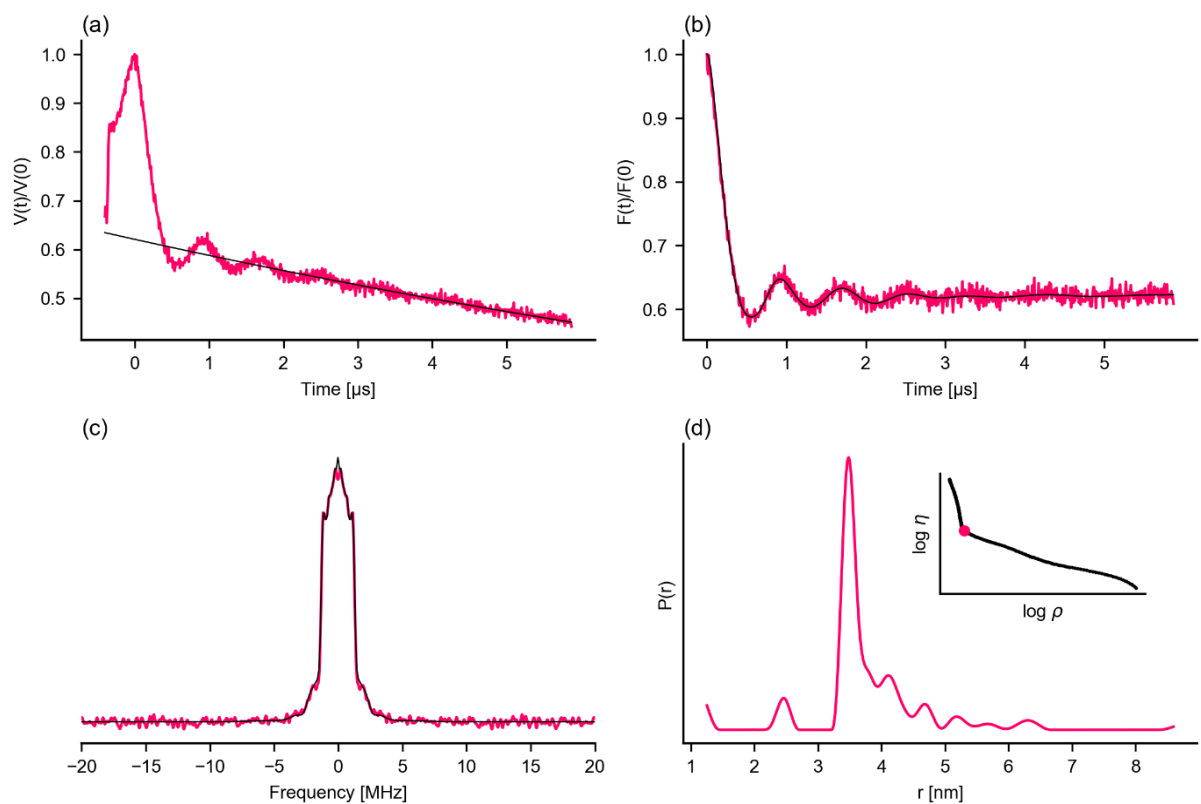

**Figure S11** Distance measurement with LiDEER (chirped pump pulse) of peptide **2**. (**a**) Normalized time trace with three-dimensional background fit. Zero time 400 ns, automatically determined background start 872 ns, cut-off at 5872 ns. (**b**) Background corrected form factor, modulation depth of 37.9 %, with fit by Tikhonov regularization. (**c**) Dipolar spectrum obtained by Fourier transformation and corresponding fit. (**d**) Distance distribution obtained by Tikhonov regularization, ⟨r⟩ = 3.53 nm and ⟨s⟩ = 0.14 nm. The L curve is shown as inset, the chosen α-parameter of 185.5 is indicated.

# 9    Supporting information: References